\documentstyle[11pt,aaspp4,psfig]{article}   %%% This makes "preprint" smaller
\def\Msun{\,M$_\odot$}
\def\lm{low-mass BH binaries } 
\def\ace{$\alpha_{\rm CE}$ }
\def\bh{BH binaries }
\begin{document}

%\slugcomment{Submitted to The Astrophysical Journal.} 

\title{Donor Stars in Black-Hole X-Ray Binaries}
\author{Vassiliki Kalogera}
\affil{Harvard-Smithsonian Center for Astrophysics, 60
Garden St., Cambridge, MA 02138; vkalogera@cfa.harvard.edu}

\begin{abstract}

We study theoretically the formation of black-hole (BH) X-ray binaries.
Consistency of the models with the observed relative numbers of systems
with low-mass ($\lesssim 2\,$M$_\odot$) and intermediate-mass ($\sim
2\,$M$_\odot - M_{\rm BH}$) donors leads to severe constraints on the
evolutionary parameters of the progenitors. In particular, we find that
(i) BH progenitor masses cannot exceed about 2\,$M_{\rm BH}$; (ii) high
values of the common-envelope efficiency parameter ($\alpha_{\rm CE} > 1$)
are required, implying that energy sources other than orbital contraction
must be invoked to eject the envelope; (iii) the mass-loss fraction in
helium-star winds is limited to be $\lesssim 50$\%. Outside of this
limited parameter space for progenitors we find that either BH X-ray
binary formation cannot occur at all or donors do not have the full range
of observed masses.  We discuss the implications of these results for the
structure of massive hydrogen-rich stars, the evolution of helium-stars,
and BH formation. We also consider the possible importance of asymmetric
kicks.

\end{abstract}


\section{INTRODUCTION}

Radial velocity measurements of the non-degenerate donors in X-ray
binaries provide limits on the masses of the accreting objects. When
combined with additional information (e.g. spectra of the donor; orbital
light curves) actual measurements of the masses are obtained (for a recent
review see Charles 1998). In this way, the masses of eleven compact
objects in X-ray binaries have been found to exceed the maximum mass
possible for a neutron star (the most recent one by Orosz et al. 1998; for
earlier determinations see the review by Wijers 1996). These binaries are
thought to be black-hole (BH) X-ray binaries. 

In addition to these eleven systems, there are X-ray sources for which
radial velocity measurements have not yet been obtained but whose X-ray
spectral and variability properties show similarities with the dynamical
BH X-ray binaries (e.g., Chen, Shrader, \& Livio\markcite{C97} 1997;
Stella et al.\markcite{S94} 1994).  These sources are regarded as possible
BH candidates. We note, however, that similar spectral and variability
characteristics have also been observed in some neutron star binaries
(van~der~Klis\markcite{vdK94a}\markcite{vdK94b} 1994a,b). Indeed, some of
the systems once considered to be BH candidates have later turned out to
contain neutron stars, while others did turn out to contain black holes
(e.g., 4U 1543-47). Therefore, the presence of a black hole in these
binaries is somewhat uncertain. 

The recent increase of the number of known binaries with accreting black
holes in our Galaxy allows us to study them as a separate population of
X-ray binaries.  In particular, the subset of eight dynamically identified
BH systems with Roche-lobe filling, low-mass companions appears to form a
uniform class of sources, typically discovered as X-ray transients. 

One issue of interest concerns the total number of BH binaries in
our Galaxy and their birth frequency. Empirical estimates have been
hampered by the transient nature of the systems and the lack of good
constraints on their recurrence times from present data. Recently,
Romani\markcite{R98} (1998) considered in detail the sensitivity and sky
coverage of the main X-ray surveys. He estimated that there are $\sim
1700$ \lm in our Galaxy with a typical discovery rate of $\sim
2$\,yr$^{-1}$.  On the theoretical side, the birth rate of \lm can be
calculated using population synthesis methods. Based on such calculations
it has been pointed out that the formation of \lm can be considerably
favored relative to neutron-star binary formation if there is limited
mass loss during the collapse and if kick magnitudes are small
(Romani\markcite{R92}\markcite{R96} 1992, 1996). It has also been
suggested (Portegies Zwart, Verbunt, \& Ergma\markcite{PZ97} 1997) that
the formation of \lm at rates comparable to the, notably rather
uncertain, empirical estimates ($\sim 10^{-7}$\,yr$^{-1}$) may require
high values of the efficiency for common-envelope (CE) ejection. However,
different choices for the mass-ratio distribution in primordial binaries
and of average kick magnitude can affect the predicted birth rates by one
or two orders of magnitude (see Fryer, Burrows, \& Benz\markcite{F98}
1998;  Kalogera \& Webbink\markcite{KW98} 1998, hereafter KW98). 

Another point of interest concerns the types and masses of donors found in
the observed BH binaries. Just like neutron-star binaries, BH binaries can
be separated into two classes: one with black holes accreting from the
wind of a massive star (3 out of 11 systems) and another with low-mass
Roche-lobe filling companions.  For accreting neutron stars this division
has been understood in the context of unstable mass transfer, which occurs
approximately when the companion is more massive than the accretor (van
den Heuvel\markcite{vdH75} 1975;  Kalogera \& Webbink\markcite{KW96}
1996). The development of this instability limits the masses of Roche-lobe
filling donors to be $\lesssim 1-2$\Msun. Analogous considerations in the
case of \lm would naively lead us to expect donors as massive as the black
holes ($\sim 10$\Msun ).  However, it has been noticed that
intermediate-mass donors ($\gtrsim 2$\Msun ) to black holes appear to be
rare (Wijers 1996).  Of the eight Roche-lobe overflow systems only {\it
two}, J1655-40 and 4U1543-47, are thought to have donor masses of $\sim
2.3$\Msun\  and $\sim 2-3$\Msun , respectively; the other six have masses
well below 1\Msun\  (e.g., Charles 1998).  Of the systems thought to harbor
a black hole based on their spectral and variability properties, there are
six more sources for which we have some information about their visual
magnitudes, $m_V$.  Measurements or just lower limits on $m_V$ combined
with estimated distances and extinction strongly indicate that the donors
have masses $\lesssim 1$\Msun\  (Ritter \& Kolb\markcite{RK98} 1998; Chen et
al.\ 1997; Stella et al.\ 1994). 

In this paper we address the issue of the apparent paucity of
intermediate-mass donors in \lm by considering a larger set of
constraints, beyond the stability of mass transfer, imposed both on
BH binaries with donors filling their Roche lobe on the main
sequence (MS) and on their progenitors.  In the present study we assume
that no asymmetric kicks are imparted to black holes at the time of their
formation. This is consistent with our limited understanding of the
physical mechanism for supernova kicks and for BH formation.
Under this assumption, we find that formation of BH binaries is
actually not possible for a wide range of evolutionary parameters.  When
possible, it is often the case that systems are formed either with only
low- or only intermediate-mass donors. This ``dichotomy'' has its origin
mainly in the fact that magnetic braking operates only for low-mass
stars. Furthermore, we find that in cases where formation of both types
of donors is possible, consistency with observations requires that the
efficiency for CE ejection be relatively high (in agreement
with the results of Portegies Zwart et al.\ 1997, although based on
different considerations). We also show that the progenitors of black
holes can be at most about twice as massive as the black hole being
formed. Finally, we are able to place very strict limits on the wind mass
loss from relatively massive helium stars. For the formation of black
hole binaries with main sequence companions to be {\em at all} possible,
the amount of mass lost from helium stars must be smaller than about 50\%
of their initial mass. 

In the next section we discuss the BH formation path considered here and
describe the evolutionary constraints that are imposed on the systems and
their progenitors. In \S\,3 we calculate the corresponding limits set on
the binary parameters. In \S\,4, we present our results on the masses of
donors in \lm that are allowed to form when all the constraints are taken
into account. We also investigate their dependence on various evolutionary
parameters, varied in the entire permitted range.  In \S\,5, we compare
the lifetimes and predicted birth rates for the two groups of systems
(with low- and intermediate-mass donors).  We find that the results are in
consistent with the observations if certain evolutionary parameters are
restricted within relatively narrow ranges. Finally, in \S\,6, we discuss
the implications of our results for our understanding of the formation of
BH X-ray binaries, the possible selection effects acting on the observed
sample, the viability of other evolutionary paths, and the relevance of
kicks possibly imparted to black holes at birth.

\section{FORMATION OF BLACK-HOLE BINARIES POWERED BY ROCHE-LOBE OVERFLOW}

\subsection{Formation Channel} 

Over the years, one evolutionary sequence has developed as being the
standard formation path for low-mass X-ray binaries with neutron stars
(Sutantyo\markcite{S75} 1975; van den Heuvel\markcite{vdH83} 1983).  It
involves the evolution of binaries with extreme mass ratios ($\lesssim
0.1$) through a CE phase and the subsequent collapse of a helium star (the
post-CE primary) into a neutron star. The X-ray phase is initiated when
the low-mass companion to the compact object fills its Roche lobe, either
because of its own radial expansion on the giant branch or because of
orbital angular momentum losses due to gravitational radiation and
magnetic braking. Naturally, this evolutionary channel can be extended to
account for the formation of \lm starting from primordial binaries with
presumably more massive primaries so that the helium star collapses into a
black hole instead of a neutron star. This evolutionary sequence has
already been considered for the formation of \lm in earlier studies
examining the birth rate of such systems or the masses of the BH
progenitors (Romani 1996; Portegies Zwart et al.\ 1997; Ergma \& van den
Heuvel\markcite{E98} 1998). 

Another channel has also been proposed that involves the evolution of a
hierarchical triple system (Eggleton \& Verbunt\markcite{E86} 1986),
which consists of an inner massive binary and a third, distant low-mass
companion. This channel has not been so far studied in any quantitative
way. However, in recent years, multiple stellar systems have been
discovered at a significant rate (for a recent catalogue see
Tokovinin\markcite{T97} 1997) and detailed modeling of triple-star
evolution could prove quite interesting. We note, though, that a crucial
phase in this sequence is the formation of a Thorne-Zytkow object
(massive envelope accreting on a compact object) and its subsequent
evolution. The stability and fate of such a configuration is still an
issue of debate (Biehle\markcite{B91} 1991; Cannon et al.\markcite{C92}
1992; Fryer, Benz, \& Herant\markcite{F96} 1996). 

One issue that is common to both formation paths is that progenitor
systems with extreme mass ratios are required, since the donor star in
the low-mass BH binary is much less massive than the progenitor
(massive single star or inner binary system) of the black hole. Among the
observed binaries in our Galaxy most have stellar components of
comparable mass, a smaller fraction have mass ratios (secondary to
primary mass) as low as $\sim 0.2-0.3$, and no systems have been observed
with mass ratios $\lesssim 0.1$, primarily because such systems are
inaccessible to current instrumentation. It has been shown, however, that
this kind of distribution is only a result of strong selection effects
favoring systems with stars of comparable brightness, hence mass
(Hogeveen\markcite{H91} 1991). Nevertheless, even correcting for the
selection effects cannot provide us with any information about the
existence or the true relative frequency of systems with extreme mass
ratios ($\lesssim 0.1$). 

If systems with extreme mass ratios do not exist in nature, then there is
no alternative channel at present for the formation of \lm. This is in
contrast to low-mass X-ray binaries (LMXBs) with neutron stars, which, in
the absence of very-low mass-ratio binaries, could still all be formed by
accretion induced collapse (AIC) of accreting white dwarfs. The
possibility of neutron star formation via AIC is still highly
controversial. However, accretion-induced collapse of neutron stars cannot
possibly account for the observed BH binaries since the measured BH masses
exceed the sum of those of neutron stars and their low-mass companions
(necessary for stable, long-term, and -- in the most favorable case --
conservative mass transfer).  One way to avoid the requirement of extreme
mass ratios would be for the black holes to acquire their low-mass
companions after their formation, but this is extremely unlikely in the
low-density environment of the Galactic disk; ejection of systems from
globular clusters also appears unlikely given the small vertical scale
height inferred for the population (White \& van~Paradijs\markcite{W96}
1996; however see Mardling 1996\markcite{M96}). Therefore, it appears that
parent multiple systems consisting of stars massive enough to produce
$\sim 10$\,\Msun\  black holes and of companions less massive than the black
hole are required for the formation of the observed \lm.

In what follows, we study the formation of BH X-ray binaries with
Roche-lobe filling donors on the main sequence via the channel that
involves CE evolution and the collapse of a helium star,
where the existence of binaries with low mass ratios is assumed, although
their relative birth frequency is not strictly specified. In the
Discussion section we also consider the fate of those primordial binaries
which are wide enough to avoid evolution through a CE phase.

\subsection{Evolutionary Constraints}

As the evolution of a binary system can follow a wide variety of paths, it
is reasonable to expect that evolution through a specific channel 
requires that binaries satisfy a given set of constraints. The complete set
of constraints relevant to the He-star collapse channel for neutron-star
binaries has been discussed in detail by KW98. We also list them here in
favor of completeness in the case of BH binaries (see also
Portegies Zwart et al.\ 1998).

Before the formation of the black hole :  
\begin{enumerate} 
\item{The binary orbit must be small enough for the massive star to fill
its Roche lobe and the system to enter a CE phase. This sets
an upper limit on the initial orbital separation of the binary. }
\item{The helium-rich primary and its companion must fit within their
Roche lobes at the end of the CE phase, for the phase to
actually be terminated. This constraint set lower limits on the orbital
separation of the binary after the CE phase. In 
addition, the helium star must fit inside the Roche lobe even
after it experiences wind mass loss as it evolves towards collapse, so
that it avoids complete merging with its companion or losing its envelope
(in most cases the carbon core will not be massive enough to collapse to
a black hole). 
For the formation of black holes of several solar masses relatively 
massive helium stars are required (the lower limit on their mass being 
set by the mass of the black hole). Calculations of helium-star 
evolution (Habets\markcite{H85} 1985; Woosley, Langer, \& 
Weaver\markcite{W95} 1995) indicate that such massive helium stars do 
not become giants at the end of their evolution. Therefore 
the requirement that
they fit in their Roche lobes turns out not to be a significant
constraint.} 
\end{enumerate}

In the last few years, the evidence for kicks being imparted to neutron
stars at birth has greatly increased (Kaspi et al.\markcite{Kas96} 1996; 
Hansen \& Phinney\markcite{H97} 1997; Lorimer, Bailes, \&
Manchester\markcite{L97} 1997; Fryer \& Kalogera\markcite{F97}; Fryer,
Burrows, \& Benz 1998). Depending on the physical origin of the kicks and
on the way black holes are formed, kicks may or may not be imparted to
black holes. Based on the spatial distribution and kinematic properties
of BH binaries, White \& van Paradijs (1996) conclude that no
asymmetric kicks are imparted to black holes. In any case, even if kicks
are associated with BH formation, they are probably smaller in
magnitude than the neutron-star kicks and possibly inversely proportional
to their mass (for a given amount momentum transfer). In what follows, we
assume that kicks are not imparted to black holes, although we discuss
their possible effect in \S\,6.  The constraints imposed on the
progenitors of BH binaries after the formation of the black hole are then
as follows: 
\begin{enumerate} 
\setcounter{enumi}{2}
\item{The system must remain bound after the collapse. For a symmetric
collapse, an upper limit is set on the amount of mass lost during 
BH formation and hence on the mass of the collapsing helium star.}
\item{For the formation of BH binaries with Roche-lobe filling donors on
the main sequence, the companion to the black hole must fill its Roche
lobe before it evolves away from the main sequence. For donors less
massive than $\sim 1$\Msun , there is an additional {\em time} constraint
that the post-collapse binary be tight enough for Roche-lobe overflow to
occur within the age of the Galactic disk (assumed here to be
$10^{10}$\,yr). These constraints set an upper limit on the post-collapse
orbital separation.}
\item{Mass transfer must proceed stably -- the donor must remain in
hydrostatic and thermal equilibrium -- and at sub-Eddington rates. This
sets
an upper limit on the mass of the donor when it is on the zero-age main
sequence, and an upper limit on the orbital separation when the mass of
the donor is comparable to that of the black hole and the star has
evolved away from the zero-age main sequence.}
\end{enumerate}

\section{LIMITS ON THE PROGENITORS}

The constraints discussed in \S\,2.2 impose specific limits on the
progenitor characteristics at different stages of their evolution. Based
on the effect of each evolutionary phase on binary parameters we
translate all of the constraints to a given stage in the formation path,
which we choose here to be the one of circularized post-collapse orbits. 

For a specific value of the mass, $M_{\rm BH}$, of the black hole, we
calculate all the limits imposed on the circularized post-collapse orbital
separations, $A$, and the donor masses, $M_d$. The exact position of the
limits on the $A-M_d$ space depends on the BH mass and on three
additional quantities related to the pre-collapse evolution of the
progenitors: 

(i) {\em The common-envelope efficiency, $\alpha_{\rm CE}$. } Since the
details of CE evolution are still not well understood (for a
review see Iben \& Livio\markcite{I93} 1993), we adopt the usual
formulation for the orbital contraction in the CE phase as suggested by
Webbink\markcite{W84} (1984).  Accordingly, we assume that the common
envelope
becomes unbound as the low-mass companion spirals inwards and orbital
energy is dissipated in the envelope with some assumed efficiency
$\alpha_{\rm CE}$ defined by:  
\begin{equation}
\alpha_{\rm CE}\left(\frac{G~M_{\rm 
He}~M_d}{2~A_f}~-~\frac{G~M_{p}~M_d}{2~A_i}\right)~=~\frac{G~M_e~M_{p}}
{\lambda~R_p}~=~\frac{G~M_e~M_{p}}{\lambda~r_{L_p}~A_i}.
\end{equation}
Here $M_{p}$ is the mass of the primary at the onset of the CE phase
(which can be different from the initial mass, $M_1$, of the primary
because of wind mass loss), $M_e$ is the mass of the envelope and $M_{\rm
He}$ is the mass of the helium core of the massive primary (which becomes
the post-CE primary), $A_i$ and $A_f$ are the orbital separations at the
onset and at the end, respectively, of the CE phase, $\lambda$ is a
numerical factor of order unity that depends on the degree of the central
concentration of the primary and for evolved stars is typically assumed
to be equal to 0.5, $R_p$ is the radius of the primary at the onset of
the CE phase that is also the radius of the Roche lobe of the primary and
hence equal to $r_{L_p}A_i$ ($r_{L_p}$ being the dimensionless radius of
the primary's Roche-lobe in units of the initial orbital separation; we
adopt the approximation of $r_{L_p}$ as function of the mass ratio as
given by Eggleton\markcite{E83} 1983). 

Although, intuitively, one would expect that the value of $\alpha_{\rm
CE}$ is restricted to be between zero and unity, it has been realized
that additional sources of energy, other than orbital, may contribute to
the ejection of the envelope; these include the thermal and ionization
energy of the envelope (e.g., Han, Podsiadlowski, \&
Eggleton\markcite{H95} 1995) or possibly nuclear energy generated in the
region close to the in-spiraling companion (Taam\markcite{94} 1994). In
the formulation of equation (1), such additional energy sources imply
values of $\alpha_{\rm CE}$ higher than unity. 

Scrutiny of equation (1) indicates that, for a given decrease in orbital
separation, the absolute normalization of \ace is not well defined as it
depends on the central-concentration parameter $\lambda$. The value of
$\lambda$ is quite uncertain, as the usual assumption of $\lambda=0.5$ is
based on models for solar-mass giant stars and an extrapolation is
necessary to the high stellar masses of interest here. In a similar way
(see eq.\ [1]), the definition of \ace depends on how accurately known are
the radii of massive stars typically after they have left the main
sequence ($R_p=r_{L_p}A_i$). In calculating the upper limit on $A$ for CE
evolution, the maximum stellar radius, $R_{\rm max}$, for which Roche-lobe
overflow is possible, enters.  For the massive primaries of interest here
this maximum radius is reached typically at core Helium ignition and
sensitively depends on the details of the stellar evolution models
(treatment of convection, mass-loss rates, etc.). For a given decrease in
orbital separation, deviations from the assumed values of $\lambda$ (0.5)
and $R_{\rm max}$ ($R_{\rm SCH}$ from the Schaller et al.\ [1992] models)
affect the normalization of \ace , $\alpha_{\rm CE}~(R_{\rm max}/R_{\rm
SCH})~(\lambda /0.5)$, and therefore any conclusions regarding
contributions from energy sources other than that of the orbit.

(ii) {\em The ratio of the helium-star mass at the time of its collapse,
$M_{\rm He,f}$, to that at the end of the CE phase, $M_{\rm He}$.} Bare
helium stars are known to lose mass in often strong Wolf-Rayet winds
(e.g., Barlow, Smith, Willis\markcite{B81} 1981).  Although observational
determinations of Wolf-Rayet mass-loss rates have been improved in recent
years, the dependence of these rates on fundamental stellar parameters
(e.g., mass, radius, luminosity, etc.) still remains very uncertain.
Langer\markcite{L89} (1989) proposed a mass-loss law such that the
mass-loss rate depends only on the stellar mass (power-law dependence
with an index of 2.5). Stellar evolution calculations adopting this law
(Woosley et al.\ 1995) results suggest that, for a wide range of initial
helium-star masses (up to $20$\Msun ), the final masses are concentrated
in a narrow range between 3 and 4\Msun . Such final masses are certainly
too small when compared to the BH masses ($5-10$\,\Msun ) in the observed
X-ray binaries.  Given the uncertainties in determining the mass-loss
rates and their dependence on stellar parameters, we choose to treat the
final helium-star mass as a free parameter. The limits set on the binary
parameters based on the constraints discussed in \S\,2.2 depend only on
the ratio $M_{\rm He,f}/M_{\rm He}$, which can in principle have any
value between zero and unity. 

(iii) {\em Mass ejection during the formation of the black hole}. 
Black-hole formation during the collapse of massive stars is still not
well understood. From core-collapse simulations (for a recent review, see
Burrows\markcite{B98} 1998) it has become clear that the nature (NS or
BH) and the mass of the remnant, the amount of fallback, and the amount
of mass ejection depend sensitively on the details of the collapse
mechanism, the energy involved, and the structure of the collapsing star.
Therefore, we are not yet in a position to identify the mass of the
progenitor, given a BH mass (there may not even be a unique
answer). Mass ejection can be significant both in cases where BH
formation occurs promptly and when it is preceded by the formation of a
neutron star, which then collapses to a black hole because of heavy
fallback (e.g., Woosley \& Weaver\markcite{WW95} 1995; Fryer \&
Woosley\markcite{FW98} 1998). Given our limited current understanding of
these processes, we allow for the mass of the collapsing helium star
$M_{\rm He,f}$ to exceed the mass of the black hole. 

An upper limit on $M_{\rm He,f}$ is set by the constraint that the 
post-collapse system is bound. For a symmetric collapse this implies: 
\begin{equation}
M_{\rm He,f} - M_{\rm BH} ~ \le ~ \frac{M_{\rm He,f}+M_d}{2}.
\end{equation}
In what follows we let the ratio $M_{\rm He,f}/M_{\rm BH}$ be a free 
parameter constrained in the range
\begin{equation}
1 ~ \le ~ \frac{M_{\rm He,f}}{M_{\rm BH}} ~ \le ~ 2~+~\frac{M_d}{M_{\rm 
BH}} ~ \lesssim ~ 3, 
\end{equation}
where the first inequality follows from the minimal requirement that the
mass of the collapsing helium star is at least as massive as the black
hole, and the last inequality follows from the condition that the donor
mass, $M_d$, has to be smaller than the BH mass (approximately)
for the mass transfer in the X-ray phase to remain stable and at
sub-Eddington rates. Note that whether helium stars with final masses in 
the range defined by equation (3) will actually collapse to black holes 
is quite uncertain. It is possible that there a more restrictive lower 
limit on $M_{\rm He,f}$ or that there are additional physical parameters 
(e.g., magnetic field, rotation etc.) that determine the nature of the 
remnant (Ergma \& van den Heuvel 1998; Fryer \& Woosley 1998)

For a given BH mass, the above three parameters specify uniquely
the efficiency of energy deposition in the common envelope, the mass of
the collapsing helium star, $M_{He,f}$, and the initial (post-CE) 
helium-star mass, $M_{He}$.  In the present study, we derive our results
for a wide enough range of normalized $\alpha_{\rm CE}$ values (as shown
below in \S\,3), and for the complete range of values allowed a priori
for the last two ratios, $M_{He,f}/M_{He}$ and $M_{He,f}/M_{\rm BH}$. 

To calculate the exact limits imposed by the first two constraints
(\S\,2.2) we also need information regarding the masses and radii of
hydrogen- and helium-rich, mass-losing stars.  For those we adopt the
fitting relations given by KW98 (see eq.\ [A1], [A4], [A10] in KW98), 
extended to the mass ranges of interest here: 
\begin{equation}
\log M_{1,i}~=~0.58\log M_1~+~0.62 \hspace{3.cm} \log M_1 \geq 1.551
\end{equation}
\begin{equation}
\log R_{1,i}~=~-2.43\log M_1~+~7.31 \hspace{3.cm} \log M_1 \geq 1.78
\end{equation}
\begin{equation}
\log M_{He}~=~1.68\log M_{1,i}~-~1.482 \hspace{3.cm} \log M_1 \geq 1.551,
\end{equation}
where $M_1$ is the initial mass of the primary, $M_{1,i}$ and $R_{1,i}$ are 
the mass and radius, respectively, of the primary at core-helium 
ignition, and $M_{He}$ is the initial (post-CE) mass of the helium star. 
Note that we include the effect of wind mass loss (on
time scales longer than the orbital period) from both the massive
BH progenitors and the helium stars, under the assumption that the
specific angular momentum of the wind is equal to that of the mass-losing
star (Jeans mode of mass loss). In this case, the ratio of the final to
the initial orbital separation is:
\begin{equation}
\frac{A_f}{A_i}~=~\frac{M_i+M_d}{M_f+M_d}, 
\end{equation}
i.e., the orbit expands. For the massive hydrogen-rich primaries, though,
this expansion is overtaken by the stellar expansion in radius as they
leave the main sequence; the upper limit on $A$ for CE evolution is then
determined by the maximum radius of the mass-losing star typically
reached just before core-He ignition (Schaller et al.\ 1992). 

The first of the two pre-collapse constraints of \S\,2.2 for CE 
evolution to occur requires that: 
\begin{equation}
A_{\rm pre-CE} < R_{1,i}/r_{L_1}. 
\end{equation}
Using equation (1) the upper limit on $A_{\rm post-CE}$ can be obtained. 
The phases of wind mass loss from the helium star, as well as of mass loss 
during 
BH formation and subsequent circularization also affect the orbit: 
\begin{equation}
\frac{A_{\rm pre-coll}}{A_{\rm post-CE}}~=~\frac{M_{\rm He}+M_d}
{M_{\rm He,f}+M_d}, 
\end{equation}
and 
\begin{equation}
\frac{A_{\rm circ}}{A_{\rm pre-coll}}~=~(1+e)~=~\frac{M_{\rm
He,f}+M_d}{M_{\rm BH}+M_d}, 
\end{equation}
respectively, under the assumptions of Jeans mass-loss mode and symmetric
collapse; $e$ is the post-collapse eccentricity. Using equations (9)
and (10) we obtain the upper limit (for CE evolution) on the separation
of the binary at the post-collapse, circularized stage.  The effect of
circularization is calculated assuming conservation of orbital angular
momentum. We have also evaluated the circularization timescale for stars
with radiative envelopes and, as the stars approach Roche-lobe overflow,
these timescales become short ($<10^7$\,yr) compared to the timescale of
stellar expansion or orbital contraction; for stars with convective
envelopes this timescale is even shorter (Zahn\markcite{Z77}\markcite{Z92}
1977, 1992). 

The second constraint of \S\,2.2 sets a lower limit to the post-CE 
binary separation: 
\begin{equation}
A_{\rm post-CE}~>~R_d/r_{L_d}, 
\end{equation}
where $r_{L_d}$ is the radius of the Roche lobe of the secondary $M_d$ in 
units of the orbital separation and $R_d$ is the (zero-age main 
sequence) radius of the secondary (eq.\ [A1] in Kalogera \& Webbink 
1996). We then translate the above lower limit at the 
post-collapse, circularized stage with the use of equations (9) and 
(10). 

The limits imposed by the last two constraints of \S\,2.2 are independent
of the above three parameters, and depend only on the masses of the black
hole and its companion. The upper limit set on the orbital separation
(for given $M_d$ and $M_{\rm BH}$) by the age of the Galactic
disk and the Terminal Main Sequence (TMS) also depends on the strength of
angular momentum losses mainly due to magnetic braking (gravitational
radiation is included as well). Here, we use the parametrization of
magnetic braking as derived by Rappaport, Verbunt, \& Joss\markcite{R83}
(1983) and modified by a mass-dependent efficiency estimated by KW98 to fit
the rotational data of main-sequence stars more massive than the Sun
(eq.\ [A14],[9] in KW98). The magnetic braking strength implied by this
specific law has also been found to be consistent with the fraction of
short-period low-mass X-ray binaries with neutron stars and the transient
fraction among them (Kalogera, Kolb, \& King\markcite{KK98} 1998). 

Finally, to calculate the limit imposed on the binary parameters by the
constraint that mass transfer is stable and proceeds at sub-Eddington
rates, we extend the results of Kalogera \& Webbink (1996) to more
massive accretors and hence more massive donors, making sure that this
extension is consistent with the detailed calculations by
Hjellming\markcite{H89} (1989). Note that, if we were to adopt the less
strict requirement of only hydrodynamic equilibrium for the donor and
relax the constraints for thermal equilibrium or sub-Eddington
mass-transfer rates, our conclusions are further strengthened, as the
upper limit on the donor mass increases and the formation of BH
binaries with intermediate-mass donors is more favored. 

We are now in the position to study all the limits relevant to BH
binary formation. In what follows we examine the case of X-ray binaries
forming with 7\Msun\  black holes, as an example, motivated by the results
of Bailyn et al.\markcite{Bl98} (1998); they conclude that the sample of
dynamically measured BH masses in binaries with Roche-lobe main-sequence
donors is consistent with a relatively narrow range of values around
7\Msun .  The sensitivity of our results on the assumed BH mass is
discussed in \S\,6. 

The limits from all the constraints for $\alpha_{\rm CE}=1.0$, $M_{\rm
He,f}/M_{\rm He}=1.0$ (no wind mass loss from the helium star), and
$M_{\rm He,f}/M_{\rm BH}=1.0$ (no mass ejection during the collapse) are
shown in Figure 1a. The upper and lower limits on orbital separation are
shown as thick and thin lines, respectively. For this specific case, it
is clear that {\em only} intermediate-mass ($> 2$\Msun ) donors are
allowed to form in an X-ray binary with a 7\Msun\  black hole.  The
strongly restrictive character of these limits as shown in Figure 1a
indicates that the detailed study of the constraints may provide us with
some answers to the question of the physical origin of the binary
parameters seen among the observed \lm . 

We now investigate the behavior of the limiting curves as the three
parameters, $\alpha_{\rm CE}$, $M_{\rm He,f}/M_{\rm He}$, and $M_{\rm
He,f}/M_{\rm BH}$ are varied within their full ranges of allowed values. 
Changes in the value of $\alpha_{\rm CE}$ only affect the upper limit on
$A$ (thick dotted line in Fig.\ 1; ensuring that the progenitor
experiences a CE phase), while changes in the two ratios, $M_{\rm
He,f}/M_{\rm He}$ and $M_{\rm He,f}/M_{\rm BH}$ also affect the lower
limit (thin solid line in Fig.\ 1; ensuring that the donor fits within
its Roche lobe at the end of CE evolution). The rest of the limits are
independent of changes in these three parameters.  From Figures 1a-c, we
see that as these three parameters are varied, the allowed region in the
$A-M_d$ space varies as well, so that stars of different masses and
evolutionary stages can become donors to 7\Msun\  accreting black holes.
Indeed, if we increase $\alpha_{\rm CE}$ to 2 (Figure 1b), we see that
both low- ($< 2$\Msun ) and intermediate-mass donors are now allowed. If
in addition we let $M_{\rm He,f}/M_{\rm He}=0.6$ and $M_{\rm He,f}/M_{\rm
BH}=1.3$ (Figure 1c), the lower limit (thin solid line) on $A$ moves
upwards and now excludes donors of intermediate mass. This is because the
wind mass-loss from the helium star leads to orbital expansion, and the
mass ejection during collapse leads to an increase in the post-collapse
eccentricity. 

The three cases shown in Figure 1 are merely indicative of how the
evolutionary and structural constraints that are imposed on BH
binary progenitors can affect the properties of the X-ray binaries that
are formed. In the following section we examine in more detail the
constraints imposed on the masses of the donors in BH X-ray
binaries.

\section{MASSES OF DONOR STARS}

The characteristics of BH binaries are determined by the five main
constraints discussed in \S\,2.2. Depending on the degree of orbital
contraction in the CE phase, the extent of wind mass loss from helium
stars, and the amount of mass ejection at during BH formation,
three different outcomes with regard to the donor masses are possible:
BH binaries can be formed with (i) {\em only} low-mass; (ii) {\em
only} intermediate-mass; (iii) {\em both} low- and intermediate-mass
donors. Here and throughout this paper the separation between low- and
intermediate-mass donors is chosen to be 2\Msun . 

For a specific BH mass and for a range of values of the
CE efficiency $\alpha_{\rm CE}$, we can study the types of
BH donors in the plane of the other two parameters: $M_{\rm
He,f}/M_{\rm He}$ and $M_{\rm He,f}/M_{\rm BH}$, examining the complete
range of their possible, a priori, values, $0-1$ and $1-3$, respectively. 
The results for a 7\Msun\  black hole are shown in Figure 2 for six values
of $\alpha_{\rm CE}$. The three different types of main-sequence donors
are indicated by different shadings (low-mass: light gray;
intermediate-mass: dark gray;  both:  black).  Regions of the parameter
space that are not shaded correspond to conditions for which the formation
of X-ray binaries with the assumed BH mass is {\em never}
possible.

Given the high masses of the original primaries, the contraction of the
orbits, for low values of \ace ($\lesssim 0.5$), is so strong that the
donor stars cannot fit within their Roche lobes at the end of the CE
phase. As a consequence, the binaries merge and no BH X-ray
binaries are formed. This would of course be in contrast to the
observations and therefore such low values of \ace can be excluded.  As
the degree of CE orbital contraction decreases (\ace increases), the more
massive of the donors are able to eject the massive envelopes and avoid
merging with the cores of their companions. Therefore, for $0.5 \lesssim$
\ace $\lesssim 1.0$, BH X-ray binaries with {\em only}
intermediate-mass donors are being formed. The exclusion of all low-mass
donors would still be in disagreement with the observed donor masses,
forcing us to examine the limits for even higher values of \ace .  The
formation of BH binaries with donor masses lower than 2\,M$_\odot$
becomes possible only when \ace exceeds unity. For \ace $\gtrsim
1.8-2.0$, the limit for the occurrence of CE evolution is pushed far
enough up in the $A-M_d$ plane (Figure 1) that it does not at all
interfere with the limit for the donor stars to be accommodated within
their Roche lobes at the end of the CE phase. As a result, the masses of
donors allowed for the BH binaries become independent of \ace . 
Consistency with the observed sample at a minimum level, i.e., BH
binaries with low-mass donors are allowed to form, would then require
that the normalized \ace exceed unity. 

It is also interesting to note the behavior of the donor types as the
amount of mass loss associated with helium-star winds and BH
formation varies. Both of these parameters primarily affect the way the
lower limit, related to the donor fitting within its Roche lobe at the
end of the CE phase, is translated through the subsequent evolutionary
stages of wind mass loss, symmetric black hole formation and
circularization of the orbits as the onset of mass transfer is
approached. The larger the mass loss is, the more the orbits expand,
either adiabatically from the helium-star wind, or non-adiabatically
following the collapse to a black hole. Overall, we can identify the
following systematic behavior (in the range \ace $\gtrsim 1.8-2.0$, where
the precise value of \ace becomes irrelevant). If mass loss is absent or
small, then both low- and intermediate-mass donors can be feeding
7\,M$_\odot$ black holes in X-ray binaries. For moderate mass loss,
low-mass donors {\em only} are allowed, as binaries with $M_d \gtrsim
2$\,M$_\odot$ expand so much that mass transfer could not be initiated
while the donors are on the main sequence. For low-mass donors, magnetic
braking is expected to be efficient and the associated orbital
contraction brings them at Roche-lobe overflow.  Very strong helium-star
winds that decrease the helium-star mass by more than a factor of about
three ($M_{He,f}/M_{He} \lesssim 0.35$), prevent BH binary
formation altogether, because the expansion during the wind phase leads
to orbits too wide for BH binaries with main sequence donors of
{\em any} mass to be formed. Mass loss following BH formation
also leads to orbital expansion, but it is further constrained to be at
most comparable to the black hole mass ($M_{He,f}/M_{\rm BH} \lesssim 2$)
for the binary to bound after the collapse. Therefore, mass loss in both
phases has to be sufficiently small for \bh like the observed ones to
form. 

It is interesting to examine in some detail the effect of the evolution
of the helium star on two of the limiting curves (first two of the
constraints discussed in \S\,2.2). Mass loss from the helium-star wind
and during BH formation combined with circularization of the
post-collapse orbit result in orbital expansion described by equations
(9) and (10).  Both pre-collapse constraints are actually set at the
beginning and the end of the CE phase, therefore by combining these two
equations we find that the translation of the corresponding limits
through the helium-star wind and collapse phases becomes independent of
the mass, $M_{\rm He,f}$, of the collapsing helium star: 
\begin{equation}
\frac{A_{\rm circ}}{A_{\rm post-CE}}~=~\frac{M_{\rm He}+M_d}
{M_{\rm BH}+M_d}.
\end{equation}
Therefore, for specific BH and donor masses, the limits on the
orbital separations of {\em bound} systems for CE evolution (upper) and for
the donor fitting within its Roche lobe are expected to be independent of
$M_{\rm He,f}$. Indeed Figure 2 shows that the character (low- or
intermediate-mass) of donors
in \bh   changes along straight lines on the
parameter plane of $M_{He,f}/M_{He}$ and $M_{He,f}/M_{\rm BH}$. From the
definition of these two ratios, it becomes clear that the change of donor
types occurs along lines of constant helium-star mass, $M_{\rm He}$, at the
end of the CE phase as expected. 
 
We should note, however, that, although the limits in the $A-M_d$
parameter space are independent of $M_{\rm He,f}$, we cannot completely
eliminate from the problem the final helium-star mass, because knowledge
of it is necessary to decide whether the system will remain bound after
the collapse or not. Then, given that the pair remains bound, the orbital
separation of the binary is determined independently of the mass of the
collapsing helium star. The fact that we still need to know the value of
$M_{\rm He,f}$ is what actually enables us to set quantitative limits {\em
both} on the extent of wind mass losses from massive helium stars and on
the masses of black hole progenitors. 

We can use this simplifying independency of the limits on $M_{\rm He,f}$
and further examine how the results change in response to varying {\em
only} the initial helium-star mass $M_{\rm He}$ (and correspondingly that
of the massive initial BH progenitor), and the normalized CE
efficiency, \ace . As already discussed (\S\,3), for a given decrease in 
the orbital separation, the normalization of
\ace is somewhat arbitrary as it is sensitive to the estimate of the
binding energy of the envelope, which depends on the uncertain values of
the central-concentration parameter, $\lambda$, and the maximum stellar
radius, $R_{\rm max}$. Taking into account this lack of absolute
normalization, we identify the ranges of initial helium star masses,
$M_{\rm He}$, and of the product $\alpha_{\rm CE}(R_{max}/R_{\rm
Sch})(\lambda /0.5)$, for which formation of X-ray binaries with
7\,M$_\odot$ BH and low-mass, intermediate-mass, or a mixture of
these donors are expected to be formed (Figure 3). As it was also evident
from Figure 2, the formation of {\em any} X-ray binaries is possible only
if \ace (as normalized above) is higher than $\sim 0.5$. Binaries with
exclusively low-mass donors are formed when \ace $\gtrsim 1$ and the
progenitor helium stars are more massive than the black hole by about a
factor of two ($M_{\rm He} \gtrsim 15$\,M$_\odot$ and $M_1 \gtrsim 45$).
Formation of X-ray binaries with both low- and intermediate-mass donors
is required to explain the observations and this requirement constrains
\ace to be relatively high and the initial helium stars to be at most
twice as massive as the black hole (initial primaries in the range
$25-45$\,\Msun ).

\section{RELATIVE NUMBERS OF LOW- AND INTERMEDIATE-MASS DONORS}

As discussed in the Introduction, the majority of the observed black hole
X-ray binaries harbor low-mass ($< 1$\,M$_\odot$) donor stars and only two
of the 14 BH candidates (including the ones with only spectral and
variability evidence for the presense of a black hole) have masses between
2 and 3\,M$_\odot$. For the results of the present study to be consistent
with the current observational sample, the three parameters that enter the
problem (assuming a specific BH mass) should be constrained within
the ranges of values for which BH X-ray binaries with both low- and
intermediate-mass donors can be formed (black areas in Figures 2 and 3).

In this section, we examine the predicted relative numbers of systems
with low- and intermediate-mass donors. The ratio of their numbers
depends both on the lifetimes of the X-ray phases and on the birth rates
of systems in each group. 

\subsection{X-Ray Lifetimes}

Angular momentum losses due to magnetic braking are thought to dominate
the evolution of BH X-ray binaries with low-mass donors.  For
donor masses exceeding $\sim 1$\,M$_\odot$ magnetic braking is thought to
become unimportant as the stars lose their convective envelopes. 
Convective envelopes are necessary for the generation of magnetic fields
by dynamo processes and effective magnetic braking. Here, we have adopted
the magnetic braking law proposed by Rappaport et al.\ (1983), along with
a mass-dependent magnetic-braking efficiency, $b(M_d)$, derived to match
the rotational velocity data of more massive main sequence stars (of
spectral type F0 and F5; for more details see KW98). We define the
magnetic braking timescale as $\tau_{\rm MB} = J_{\rm orb}/\dot{J}_{\rm
orb}$, where $J_{\rm orb}$ and $\dot{J}_{\rm orb}$ are the orbital
angular momentum and its derivative, respectively (see eq.\ [A14] in
KW98).  For systems with MS donors less massive than $\sim
1.2$\,M$_\odot$, we calculate this timescale to be in the range $2\times
10^{8}-10^{9}$\,yr. 

In BH binaries with intermediate-mass donors the mass transfer is
driven by the nuclear evolution of the donor. Using the results of
Schaller et al.\markcite{S92} (1992) and adopting a fitting formula (eq.\
[A9] and [A10] in Kalogera \& Webbink 1996) for the evolution of stellar
radius with time on the main sequence, we find that, for 2-7\,M$_\odot$
stars, the main-sequence lifetimes and the corresponding timescales of
radial expansion lie in the range $10^{8}-10^{9}$\,yr. Therefore, it
appears that the lifetimes of the BH binaries with low- or
intermediate-mass donors are comparable within a factor of $\sim 2$.  As
a result, the relative numbers of low- and intermediate-mass BH
binaries depend mainly on the relative birth rates for the two groups of
systems. 

\subsection{Birth Rates}

So far we have identified the regions in the parameter space of
circularized post-collapse orbital separation, $A$, and donor mass, $M_d$,
of both low- and intermediate mass BH binaries, for the full range
of values of the relevant parameters describing the evolution of their
progenitors. The probability density with which these regions are populated
depends directly on the distribution function that describes primordial
binaries with extreme mass ratios and the effects on it of the various
evolutionary stages leading to the formation of BH X-ray binaries. 

In order to calculate the predicted birth rates for the two types of \bh ,
we need to synthesize a primordial binary population and evolve it through
the formation channel of interest to us here. Since we have assumed that
the helium-star collapse is symmetric, the population synthesis calculation
is extremely simplified and can be performed fully analytically using
Jacobian transformations of the distribution function (this method has been
described in detail in KW98), for the general case where the collapse is 
not necessarily symmetric. 

We assume that the primordial binary population is characterized by three
parameters: the mass of the primary, $M_1$, the mass ratio, $q\equiv
M_d/M_1$, and the orbital separation, $A_i$ (orbits are assumed to be
circular). We adopt a field-star initial mass function as derived by
Scalo\markcite{S86} (1986), a distribution over orbital separations that
is constant in $\log A_i$ (Abt\markcite{A83} 1983), and we parameterize
the mass-ratio distribution as a power-law, proportional to $q^{-\alpha_q}$
(for more details regarding these assumptions, see KW98). 
Since we are ultimately interested in ratios of predicted birth
rates, all normalization constants become irrelevant.  The distribution
function of the primordial binary population is\footnote{For simplicity,
we use the same symbol, $F$, for the distribution functions at all
evolutionary stages.  The individual stages can be identified based on
the notation used for the masses and orbital separations.}: 
\begin{equation}
F(\log M_1, q, \log A_i) \propto M_1^{-1.7}~q^{-\alpha_q}
\end{equation}
or
\begin{equation}
F(\log M_1, \log M_d, \log A_i) \propto M_1^{\alpha_q-2.7}~q^{1-\alpha_q}.
\end{equation}

We transform the above distribution function through all the
evolutionary phases and calculate it at the stage of
circularized post-collapse orbits. The stages we have to consider are: 
\begin{enumerate}
\item{ {\em Wind mass loss from the massive primary as it evolves in 
isolation until the CE phase}. The mass of the primary decreases (down to 
$M_{1,i}$ (eq.\ [4]) for the CE upper limit on $A$) and the orbit expands
according to the Jeans mode of 
mass loss (eq.\ [7].}
\item{ {\em Orbital contraction and loss of the 
primary's envelope in the CE phase}. The mass of the primary decreases 
to $M_{\rm He}$ (eq.\,[6]) and the post-CE orbital size depends on the 
CE efficiency and the stellar masses involved (eq.\ [1]).}
\item{ {\em Wind mass loss from 
the helium-star primary}. The final helium-star mass depends on the 
value of the assumed ratio $M_{\rm He,f}/M_{\rm He}$ and the orbit again 
expands according to the Jeans mass-loss mode (eq.\ [7]).} 
\item{ {\em Symmetric BH formation and circularization}. The 
primary mass decreases to the BH mass and the orbit expands (eq.\ 
[10]).} 
\end{enumerate}

In each of the above evolutionary stages, the ratio of the final to the
initial orbital separation depends {\em only} on the stellar masses
involved and the CE parameters and not on any of the orbital
separations themselves. This results in all the derivatives of the 
form $\partial
\log A_f / \partial \log A_i$ being equal to unity and therefore the
distribution function in $\log A$ remains unaffected by these
transformations. On
the other hand, the changes in primary mass (the donor mass
remains constant throughout the evolution) depend only on {\em constants}
that appear in the fitting relations, while for the
transformations from $M_{\rm He}$ to $M_{\rm He,f}$ the relevant
derivative is equal to unity. Given the simplicity of these steps, the
distribution of binaries over $\log M_{\rm He}$, $\log
M_d$, and $\log A$, remains the same (modulo normalization constants) as
in equation~(8):  
\begin{equation}
F(\log M_{\rm He}, \log M_d, \log A) \propto 
M_1^{\alpha_q-2.7}~M_d^{1-\alpha_q}.
\end{equation}

To calculate the ratio $BR_I/BR_L$ of the predicted birth rates for BH 
binaries with intermediate-mass donors over that of systems with
low-mass donors, we integrate the above distribution function over the 
ranges of orbital separations and donor masses that are relevant to each 
of the two groups: 
%\begin{equation}
%\frac{BR_I}{BR_L}~=~\frac{\int ^{M_d^{\rm max}}_{2\,{\rm M}_\odot}~ \left( 
%\log \frac{A_{\rm max}}{A_{\rm min}}\right)~M_d^{1-\alpha_q}~d\log M_d}
%{\int ^{2\,{\rm M}_\odot}_{M_d^{\rm min}}~ \left(     
%\log \frac{A_{\rm max}}{A_{\rm min}}\right)~M_d^{1-\alpha_q}~d\log M_d}.
%\end{equation}
\begin{equation}
BR_{I~or~L}~\propto ~\int ^{M_d^{\rm max}}_{M_d^{\rm min}}\left(    
\log \frac{A_{\rm max}}{A_{\rm min}}\right)~M_d^{1-\alpha_q}~d\log M_d.
\end{equation}
For specific values of the BH mass, CE efficiency,
and initial helium-star mass (or initial primary mass), $A_{\rm min}$ and
$A_{\rm max}$ are determined by the limits discussed in \S\,3 and they are
functions of $M_d$ only. The logarithm of their ratio that appears in
equation (16) is the result of the integration of (15) over $\log A$ at a 
given
$M_d$. For the systems with low- and intermediate-mass donors the values
of $M_d^{\rm max}$ and $M_d^{\rm min}$, respectively, are equal to
2\,M$_\odot$. Finally, the dependence on $M_1$ that appears in equation (15)
becomes irrelevant when we take the ratio of the two birth rates at a
given $M_{\rm He}$, hence initial primary mass, $M_1$. 

It is evident that two unknowns enter the calculation of the predicted
birth rates: the distributions of initial orbital separations and mass
ratios. Given the narrow ranges of orbital separations of interest (see
Figure 1) the exact form of their distribution does not affect the
results in a significant way.  However, this is not true for the
mass-ratio distribution. The primordial binaries that are relevant for
the formation of \lm have mass ratios typically smaller than $0.1$, a
range which is observationally inaccessible. This situation is identical
to that of LMXBs with neutron stars, which has been discussed in detail
by KW98. In short, the essentially arbitrary choice of the power-law
index $\alpha_q$ in the mass-ratio distribution affects significantly the
absolute normalization of the predicted birth rates (but essentially not 
the distribution of the newborn LMXBs in orbital parameters).  In 
what follows, we adopt the results of Hogeveen (1991) and extend
them to mass ratios smaller than $\sim 0.3$. His results show that the
observed mass-ratio distribution is biased towards high values, while the
corrected distribution can be described by a steep power law with index
$\alpha_q=2.7$. 

We use equation (16) and the limits on orbital separations as calculated
in \S\,3 and evaluate the birth-rate ratio $BR_I/BR_L$ for $\alpha_q=2.7$
and a 7\,M$_\odot$ black hole. We plot the results against the initial
(post-CE) helium-star mass for three values of the CE
efficiency in Figure 4. For smaller and higher values of \ace, the
behavior of the ratio is very similar to what is shown for \ace =1.6 and
3, respectively. It is clear that in the parameter space in which both
types of BH X-ray binaries are allowed to form (black areas in
Figures 2, 3), there are only small regions where systems with low-mass
donors dominate the population (i.e., the ratio $BR_I/BR_L$ is smaller
than unity). At first, this result may appear counter-intuitive, given
the choice of $\alpha_q=2.7$, which strongly favors very small initial
mass ratios and hence systems with {\em low}-mass companions. However,
there is a counteracting effect: the width of the range of orbital
separations allowed to be populated by the two types of systems. Scrutiny
of Figure 1 indicates that this range is considerably wider for
intermediate-mass donors than for low-mass donors. The origin of this
difference lies in the steep slopes of the radius-mass relations of both
evolved massive stars and of the low-mass zero-age main-sequence stars.
Consistency with the observations appears to be possible only if the
CE efficiency as normalized in equation (1) is relatively
high (higher than about $2$).  Given the strong peak of the assumed
mass-ratio distribution at low values, the results shown in Figure 4
should be regarded as lower limits.  We have calculated the same
birth-rate ratios for $\alpha_q=0$ (flat mass-ratio distribution, which
is often used in the literature) and found that the ratios are increased,
i.e., intermediate-mass donors are favored even more strongly. This is
expected, given the decrease of $\alpha_q$, and contrasts even more with
the properties of the observed sample. 

From the observed sample without any corrections and for the roughly
comparable lifetime estimates for the two groups (\S\,5.1), a value of
$\simeq 2/14 \sim 10^{-1}$ is implied for $BR_I/BR_L$.  Given the
uncertainties and possible selection effects acting on the population of
BH binaries, we choose to be relatively conservative and regard
model results as being consistent with observations when the ratio of
birth rates, $BR_I/BR_L$, is lower than unity (since the lifetimes of the
two groups have been found to be roughly comparable).  For normalized
\ace values in the range $1.5-2.5$, this can be obtained only for narrow
ranges of initial helium star masses, either very close to the BH
mass (7\,M$_\odot$) or between $\sim 11-14$\,M$_\odot$ (Figure 4). For
smaller values (\ace$\lesssim 1.5$), $BR_I/BR_L$ always exceeds unity,
while for higher values (\ace$\gtrsim 2.5$), the full range ($\sim
7-15$\,M$_\odot$) of $M_{\rm He}$ values (initial primary masses in the
range $\sim 25-45$) could be consistent with the observed sample. We note
that for a flat mass ratio distribution ($\alpha_q=0$), masses are
restricted in the narrow range $11-14$\,\Msun , even for high \ace values
(\ace $>2$).

\section{SUMMARY AND DISCUSSION}

We find that the models for BH formation are consistent with the
properties of the observed sample if (i) wind mass-loss from helium stars
is limited so that they lose less than half of their initial mass (ii)
helium stars that form black holes are at most about a factor of 2 more
massive than the black holes, and (iii) CE efficiencies are relatively
high and, depending on the exact radii of massive stars and their density
profiles, significant contributions from energy sources other than the
orbit may be required. 

Our results show weak dependence on the assumed BH mass. 
Although the results presented in this paper are all for $M_{\rm
BH}=7$\Msun , we have also examined models for $M_{\rm
BH}=5$\Msun\   and found that only the shape of the boundaries in the
parameter space between the different donors is somewhat changed.
However, both the qualitative conclusions and quantitative constraints on
\ace, helium-star wind mass loss, and BH progenitors remain
essentially unaltered. 

Based on the approximate formulation of the common-envelope phase
(equation 1) we stress that the CE efficiency, \ace, lacks absolute
normalization and only relative limits can be set on it.  With this
scaling in mind, it is worth noting that an investigation of the birth
rate of \lm by Portegies~Zwart et al.\ (1997) also pointed to a need for
relatively high values of \ace. On the other hand, a study of the
properties (orbital periods and transient character) of LMXBs with neutron
stars (Kalogera et al.\ 1998) favors relatively low values of \ace ($\sim
0.5$ with the formulation used here).  This difference could possibly be
attributed to the different range of primary masses that are relevant in
the formation of the two types of LMXBs ($M_1 < 25$\Msun\  for neutron
stars) and therefore to the differences in the structure and energetics of
their massive envelopes. 

For primordial binaries with extreme mass ratios like those needed for the
formation of BH binaries, a CE phase is inevitable once the primary fills
its Roche lobe. Based on our results formation of a black hole in the
range of masses observed is then possible {\em only} if the helium star is
able to retain a significant amount of its mass -- {\em at least} half of
it (see also Ergma \& van den Heuvel 1998). This provides a new constraint
on the mass-loss process in helium stars; it suggests that the
parameterization of the mass-loss rate adopted so far (Langer
1989\markcite{L89}), which leads to a very narrow range of final masses
at $\sim 4$\,M$_\odot$ (Woosley et al.\ 1995) may not apply to stars with
relatively high initial masses.

Formation of \lm through helium-star collapse is expected once the
primordial binary experiences a CE phase. There is, however, the
possibility that the initial binary is so wide that the primary never
fills its Roche lobe during its entire lifetime and collapses directly to
a black hole. If the primary is more massive than $\sim 40$\,\Msun , then
it is possible that the star loses its hydrogen-rich envelope in a wind
(Schaller et al.\ 1992) and the helium core is once again revealed. In
this case, the problems discussed above are again relevant and we would
have to conclude that helium-star winds cannot be as strong as thought
previously. If the star is able to retain part of its hydrogen envelope,
then its final mass is much larger than the observed BH masses;
subsequent mass-loss at the time of the collapse should lead to
disruption of the binary, had the collapse been symmetric. Even if the
final mass is low enough that the mass-loss at BH formation is
not disruptive (does not exceed half of the pre-collapse total mass),
using the radius-mass relation for such massive stars (Schaller et al.\
1992), we find that the pre-collapse orbital separation is so wide ($\sim
1000$\,R$_\odot$) that the post-collapse system will never reach a
Roche-lobe overflow phase and appear as an X-ray binary (Romani 1992;
Portegies Zwart et al.\ 1997). 

Based on the above discussion it becomes evident that a system could avoid
the CE phase and still form a low-mass BH binary {\em only} if significant
kicks are associated with the formation of black holes. However, for such
systems to avoid disruption due to severe mass loss and at the same time
decrease their orbital separations to reach contact, the average kick
magnitude must be at least comparable to or higher than the typical
relative orbital velocity in the pre-collapse system
(Kalogera\markcite{K96} 1996), i.e., higher than $\sim 100$\,km\,s$^{-1}$. 
Given the mass difference between black holes and neutron stars, these
kick magnitudes would require a momentum transfer several times larger in
the case of black holes. Although this cannot be excluded (since the
physical origin of kicks is not known), it imposes significant
quantitative constraints on the process. In the case of CE evolution, the
helium-star binaries are close, with orbital separations $\sim
10-20$\,R$_\odot$ and orbital velocities in excess of $\sim
$300\,km\,s$^{-1}$, up to $\sim 900$\,km\,s$^{-1}$. As above, we cannot
exclude that kicks could affect our conclusions but then the kick
mechanism should be such that the amount of momentum transfer is higher
for black holes than for neutron stars, by a factor comparable to their
mass ratio ($\sim 10$). 

So far we have implicitly assumed that the observed sample of \lm is
representative of the total underlying Galactic population, at least from
the point of view of their donor masses. However, the apparent paucity of
systems with intermediate-mass donors could be merely the result of
selection effects. We note that accurate dynamical measurements of the
compact-object masses are possible {\em only} for soft X-ray transients,
because detailed observations of the low-mass secondaries can be
undertaken only in quiescence (e.g., Charles 1998). Therefore, if BH
binaries with companion masses in excess of 2\Msun\  were persistent
sources, then a significant population of these systems could exist in
the Galaxy and remain undetected. However, if this were the case and to
the extent that emission processes were similar, then there should be
some evidence of such a population among the X-ray binaries that are
thought to be BH candidates based only on their spectral and
rapid variability characteristics. As already discussed, all the systems
(the sample includes transient and persistent sources) for which there
are at least lower limits on their visual magnitudes appear to harbor
low-mass companions (Ritter \& Kolb\markcite{RK98} 1998; Chen et al.\
1997; Stella et al.\ 1994). Furthermore, one can estimate the mass
transfer rates expected for intermediate-mass BH binaries, using
the radius expansion rates from nuclear evolution on the main sequence.
For a substantial range of masses (up to $\sim 4$\Msun ), these are found
to be low enough for the disk instability to develop and therefore such
systems should appear as soft X-ray transients just like their low-mass
counterparts (Kolb\markcite{K98} 1998).  The apparent paucity of
intermediate-mass donors appears therefore to be real. 

In the present study, we have not addressed the issue of formation of
BH binaries with Roche-lobe filling donors on the giant branch
(V404 Cyg is the only example in the observed sample).  Our results on the 
limits imposed on orbital
separations (Figure 1) indicate that the upper limit (for CE evolution)
is so strong that it may be hard for binaries to ever form with wide
enough orbits. Such orbits would be required for the systems to evolve to 
longer periods with evolved donors as 
described by Pylyser \& Savonije (1988\markcite{P88}). 
Another way to form such systems could possibly be through
secular evolution of intermediate-mass BH binaries. As magnetic
braking is not expected to be efficient for these higher-mass stars,
orbital evolution is driven by nuclear expansion on the main sequence. 
Predictions of the fate of such systems in terms of their binary
parameters require detailed modeling of the mass transfer phase; it may
be possible that systems with low-mass donors on the giant branch form
through secular evolution and not from initially wide post-collapse
binaries. 

\acknowledgements 

It is a pleasure to thank R.\ Narayan, D.\ Psaltis, and F.\ Rasio for many
stimulating discussions and useful comments in the course of this study.  
I would also like to thank S.\ Woosley for discussions on the issue of
mass loss from helium stars, and R.\ Webbink and J.\ McClintock on
selection effects on the BH X-ray binary population, as well as U.\ Kolb,
A.\ King, A.\ Esin, K.\ Menou, and E.\ Quataert for discussions. I am
grateful to D.\ Psaltis and F.\ Rasio for a critical reading of an earlier
version of the manuscript. I would also like to thank an anonymous referee
for important comments that improved the clarity of the text.  Finally, I
am thankful for the warm hospitality of the Aspen Center of Physics, the
Astronomy Group of the University of Leicester, and the ``Anton
Pannekoek'' Astronomical Institute at the University of Amsterdam.  
Support by the Smithsonian Astrophysical Observatory through a
Harvard-Smithsonian Center for Astrophysics Postdoctoral Fellowship is
also acknowledged.

\newpage

\centerline{\psfig{figure=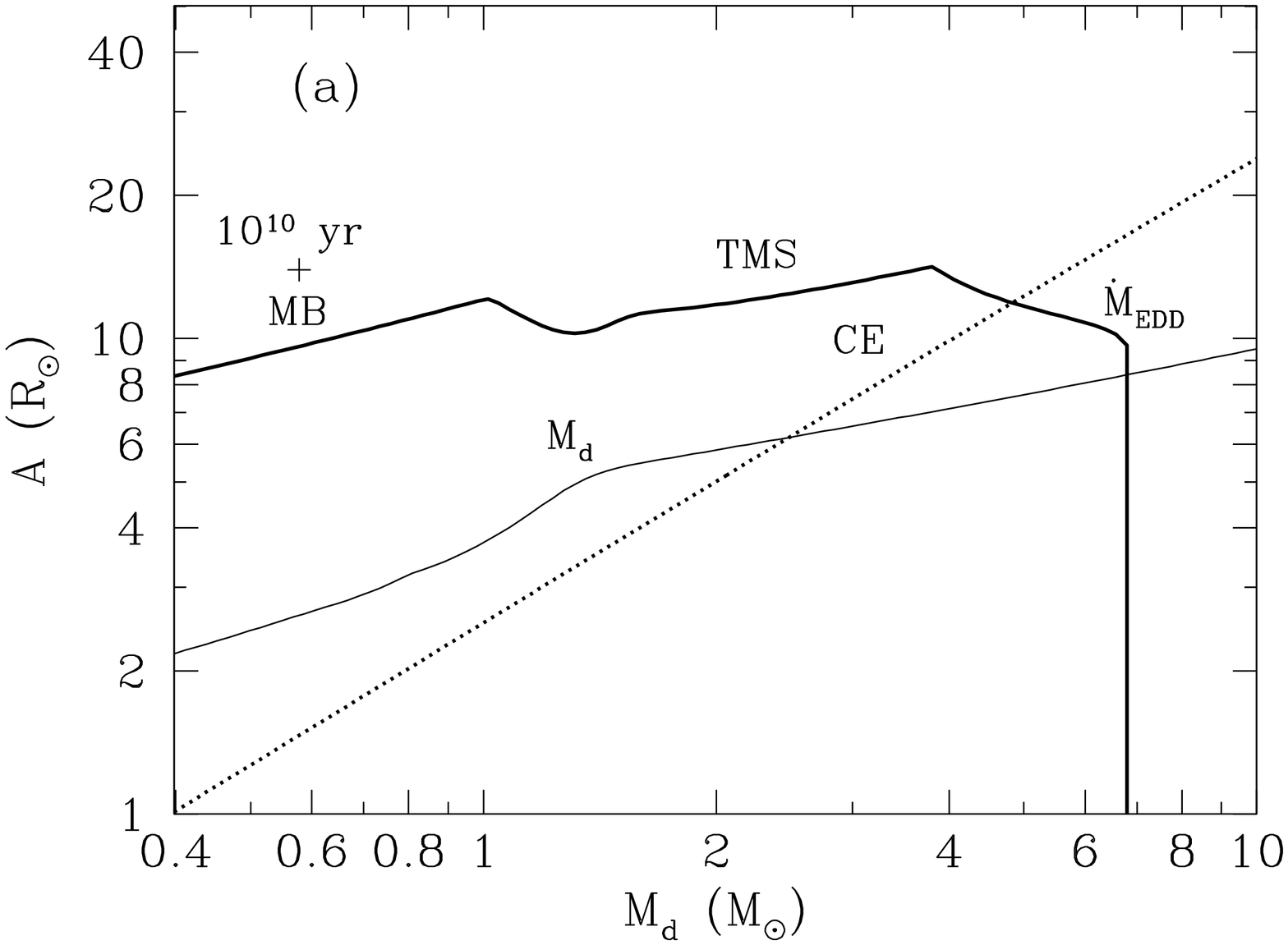,angle=-90,width=6.5in}}

\noindent
Fig.\,1 ---
Limits on the donor masses, $M_d$, and post-collapse
circularized orbital separations, $A$, of BH X-ray binaries with
a 7\Msun\  black hole. {\em Thick solid line}: Upper limit on $A$ so that
low-mass ($M_d\lesssim 1.5$\Msun ) donors will fill their Roche lobes
within $10^{10}$\,yr because of magnetic braking (MB) and that more
massive donors fill their Roche lobes before reaching the Terminal Main
Sequence (TMS) and drive at most Eddington mass-transfer rates. {\em
Thick dotted line}: Upper limit on $A$ so that the progenitors experience
a common-envelope (CE) phase. {\em Thin solid line}: Lower limit on $A$
so that the companion, $M_d$, lies within its Roche lobe at the end of
the CE phase. Limits are shown for three different choices of \ace (for
$\lambda=0.5$), $M_{He,f}/M_{He}$, and $M_{He,f}/M_{BH}$: (a) 1.0, 1.0,
1.0 (only intermediate-mass donors, $M_d>2$\Msun ); (b) 2.0, 1.0, 1.0
(both low- and intermediate-mass donors); (c) 2.0, 0.6, 1.3 (only
low-mass, $M_d<2$\Msun , donors), respectively.

\newpage

\centerline{\psfig{figure=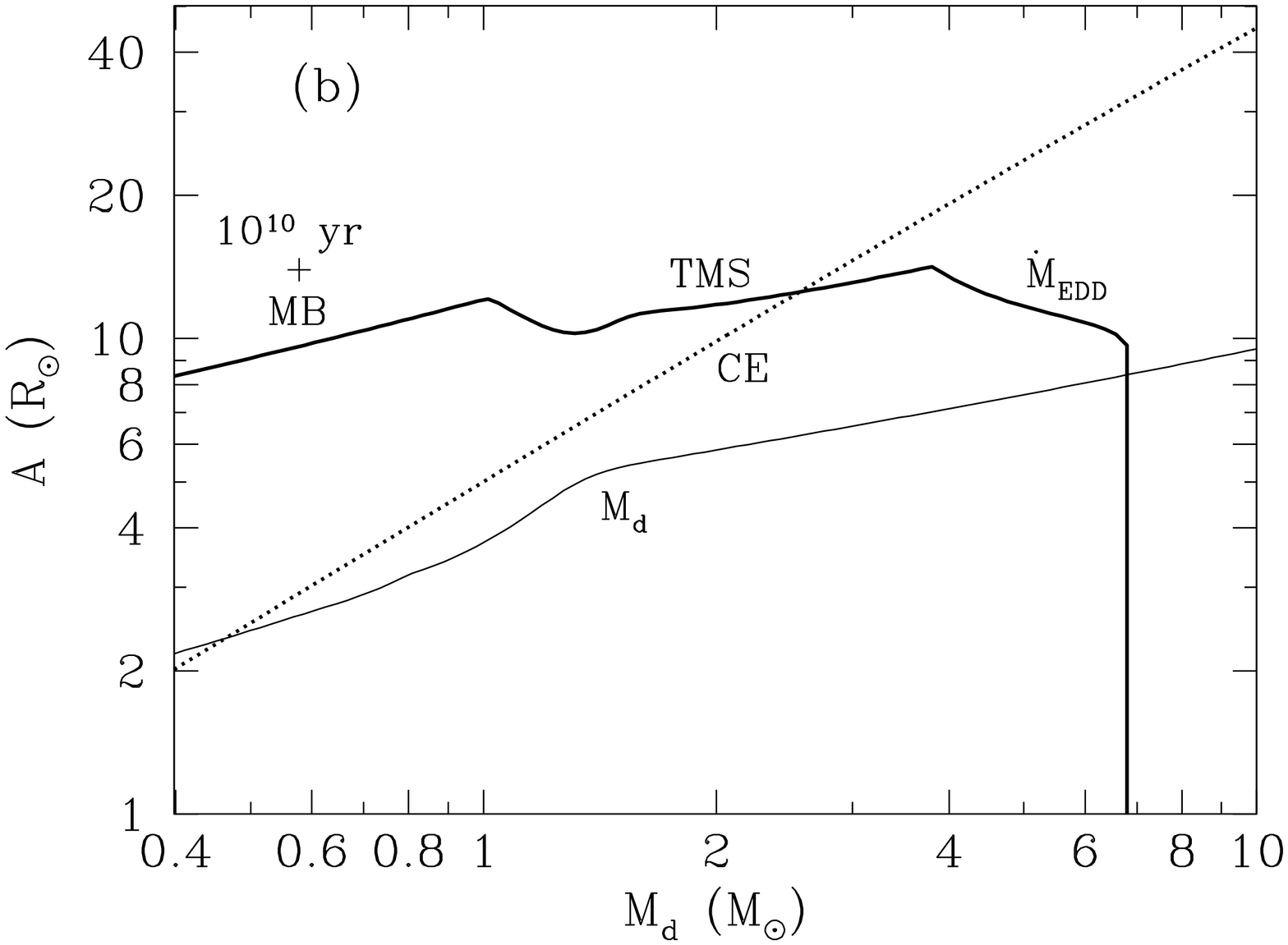,angle=-90,width=6.5in}}

\noindent
Fig.\,1b --- continued.

\newpage

\centerline{\psfig{figure=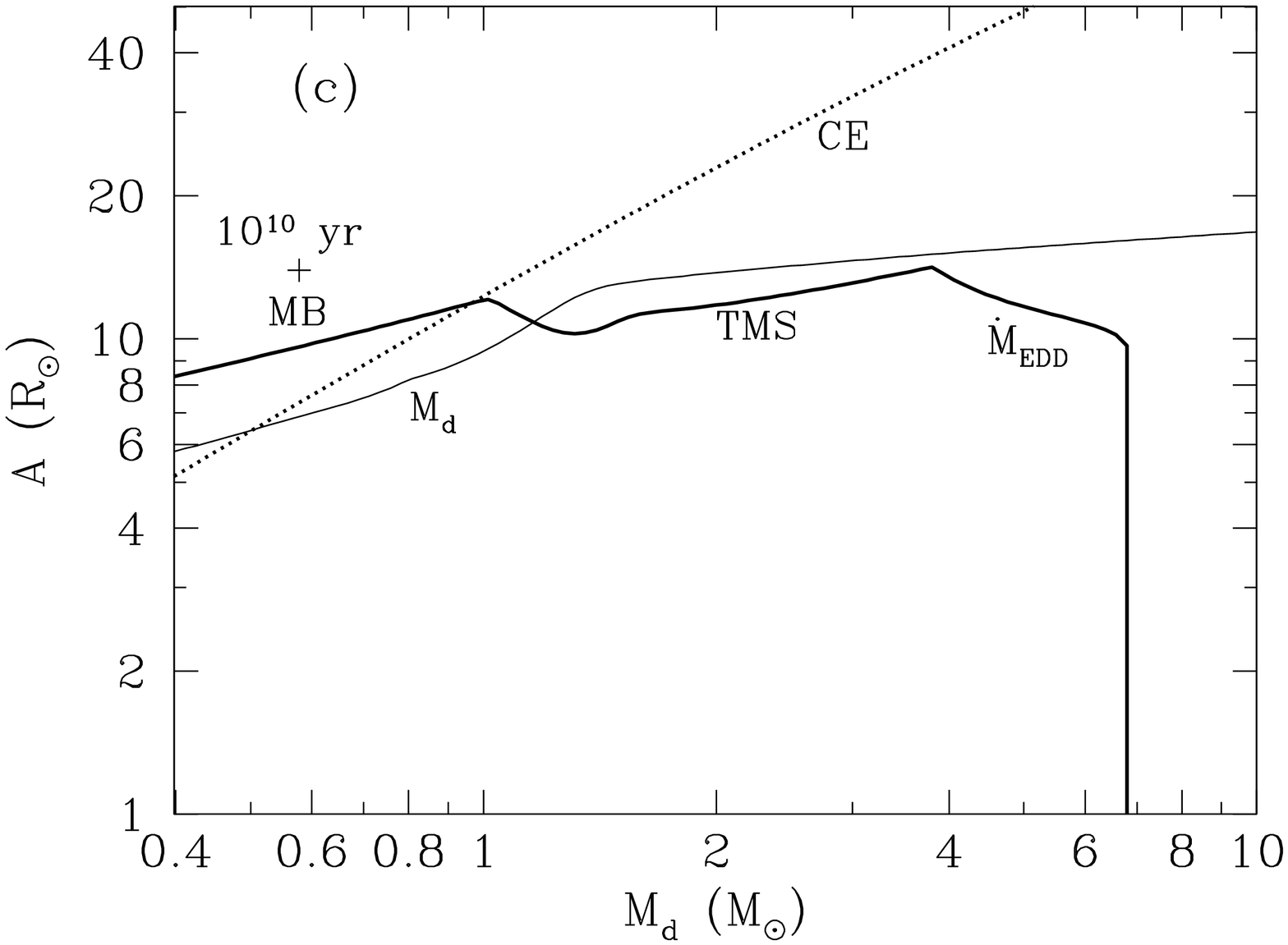,angle=-90,width=6.5in}}

\noindent
Fig.\,1c --- constinued.

\newpage

\centerline{\psfig{figure=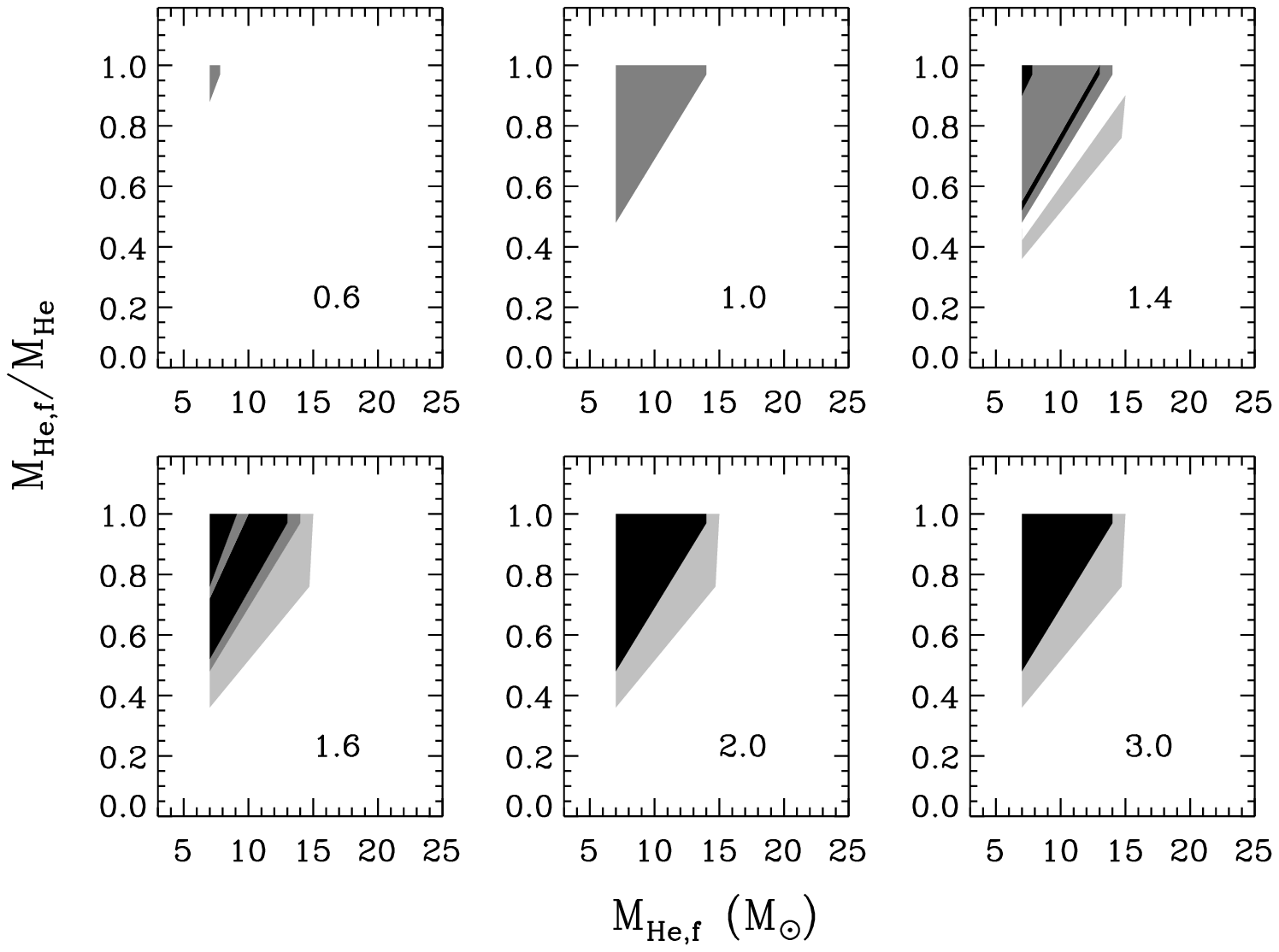,angle=90,width=6.5in}}

\noindent
Fig.\,2 ---
Limits on the parameter space of the final (pre-collapse) 
helium-star mass, $M_{He,f}$, and the ratio, $M_{He,f}/M_{He}$ that 
describes the extent of mass loss from helium stars, for six values of 
the \ace =0.6, 1.0, 1.4, 1.6, 2.0, 3.0 ($\lambda=0.5$). Conditions in 
the unshaded areas do not allow the formation of BH binaries 
with main-sequence Roche-lobe filling donors; conditions in the 
light-gray, dark-gray, and black areas allow the formation of systems 
with only low-mass, only intermediate-mass, and both types of donors, 
respectively. The BH mass has been assumed to be equal to 7\Msun . 
It is evident that relatively high values of \ace   are favored, while 
mass loss both in a helium-star wind and at BH formation are 
constrain to within a factor of about 2.

\newpage

\centerline{\psfig{figure=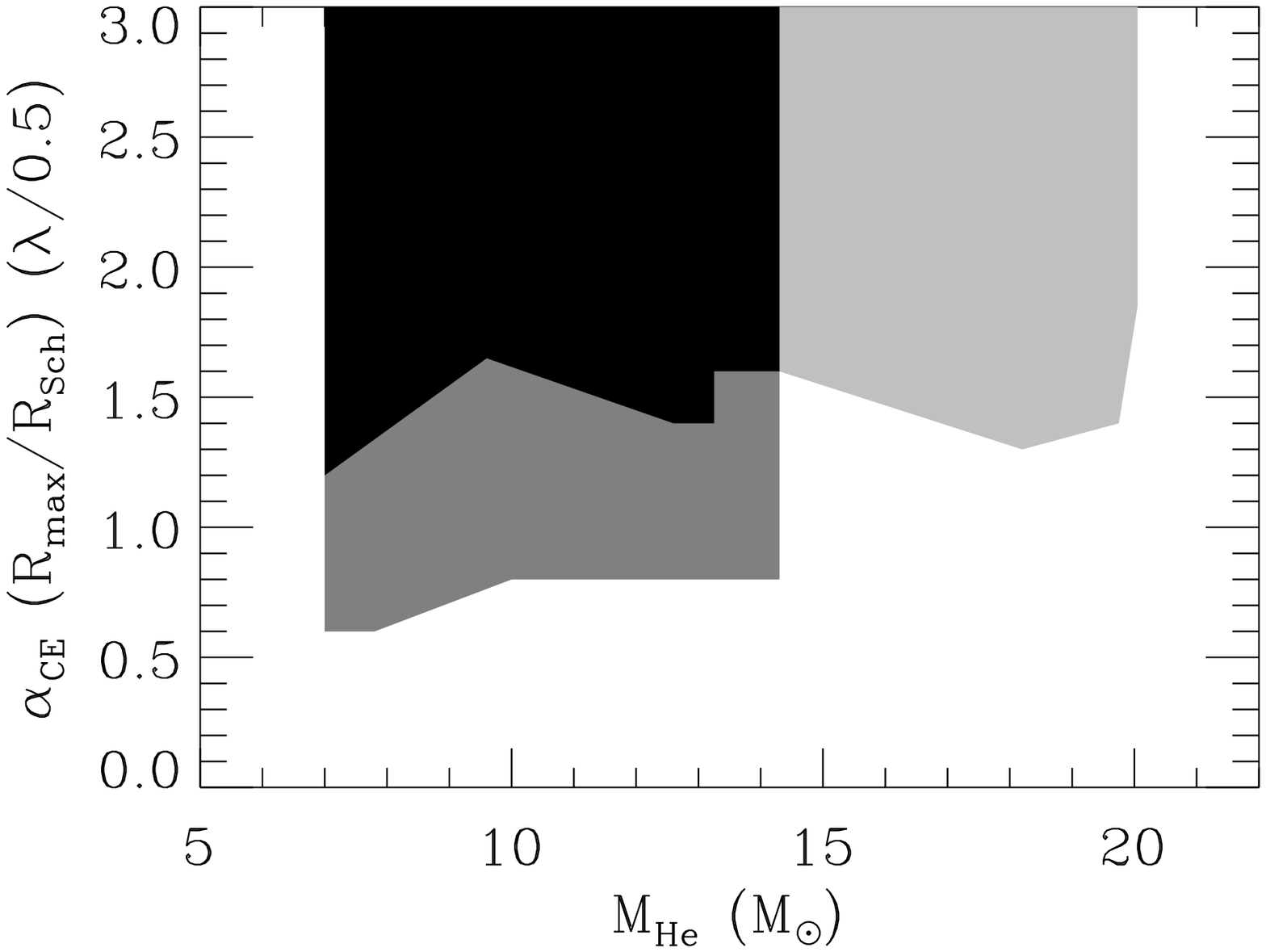,angle=90,width=6.5in}}

\noindent
Fig.\,3 ---
Limits on the parameter space of the initial (post-CE) 
helium-star mass, $M_{He}$, and the common-envelope efficiency, \ace , 
properly normalized (by the maximum stellar radii of massive stars based on 
the models by Schaller et al.\ 1992 and by the central-concentration 
parameter, $\lambda$), for a 7\Msun\   black hole. Shade coding is as in 
Figure 2. It is evident that relatively high \ace  values are 
preferred. Systems with only low-mass donors are formed if  the 
helium-star progenitors are more massive than the black hole by a factor 
in the range $2-3$, while for less massive helium stars and high \ace  
values systems with both types of donors are formed.

\newpage

\centerline{\psfig{figure=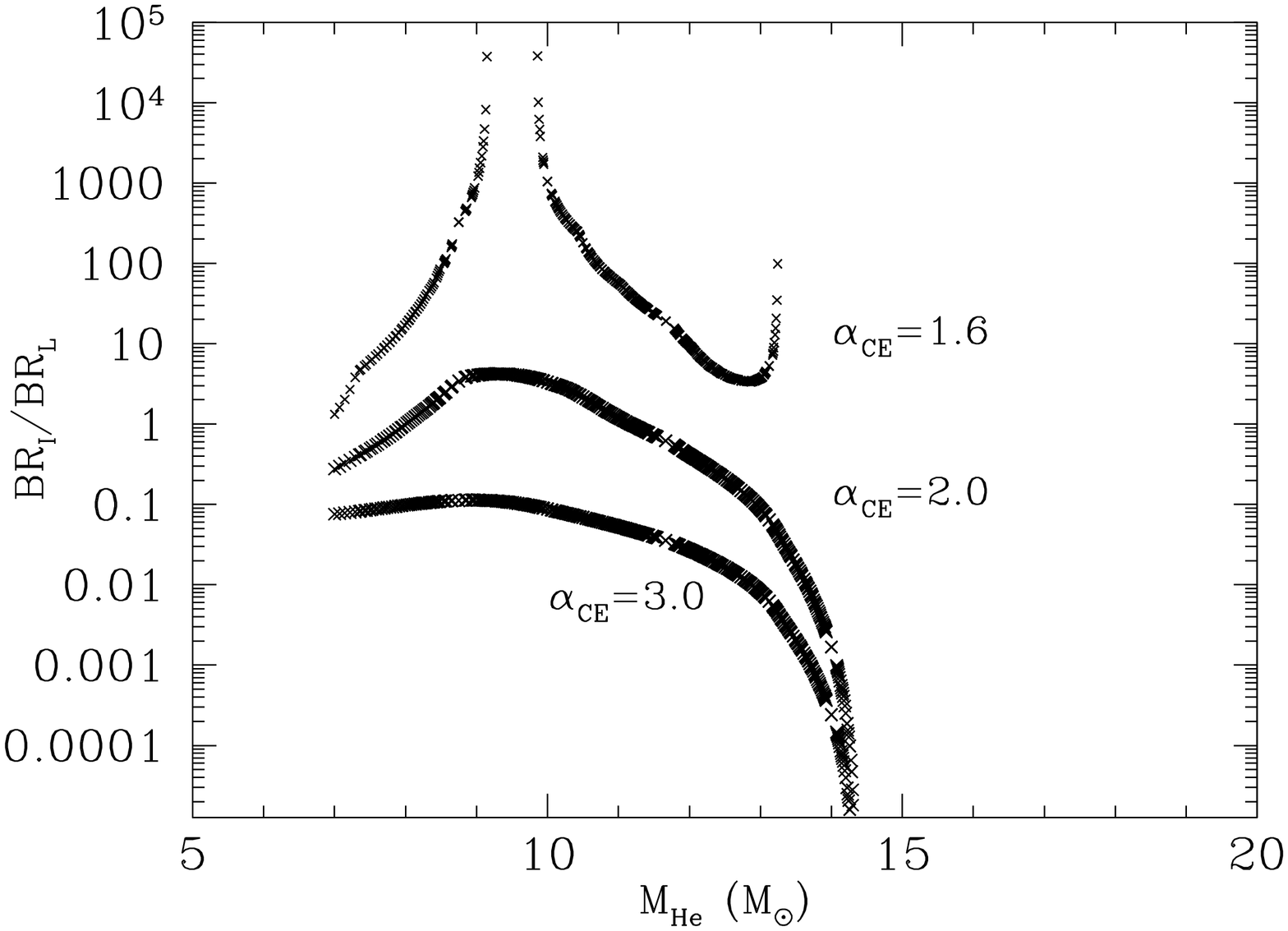,angle=-90,width=6.5in}}

\noindent
Fig.\,4 ---
Ratios of the birth rate of BH binaries with
intermediate-mass donors to that of systems with low-mass donors as
function of the initial (post-CE) helium star mass for three different
values of \ace properly normalized. The BH mass has been assumed
to be equal to 7\Msun ; the mass-ratio distribution is assumed to
strongly peak at low values and therefore the ratios shown could be
regarded as lower limits. Agreement with the observed paucity of
intermediate-mass donors seems to require moderate degrees of orbital
contraction during the CE phase, and, for non-extreme values of \ace ,
considerable mass loss during BH formation could be required.

\end{document}